\documentclass[12pt,preprint]{aastex}
\usepackage{natbib}
\usepackage{graphicx}
\usepackage{amssymb}
\usepackage{epstopdf}
\newcommand{\beq}{\begin{equation}}
\newcommand{\eeq}{\end{equation}}
\def\etal{{\sl et~al.~}}
\newcommand{\HD}{{HD~38529}}
\newcommand{\eps}{{$\epsilon$ Eri}}
\newcommand{\HST}{{\it HST~}}
\newcommand{\HIP}{{\it Hipparcos}}
\newcommand{\kms}{km s$^{-1}$~}
\newcommand{\ms}{m s$^{-1}$}

\newcommand{\msini}{$M \sin {\it i}~$}

\newcommand{\mjup}{M$_{\rm Jup}~$}
\newcommand{\mjupe}{M$_{\rm Jup}$}
\newcommand{\msun}{M$_{\odot}~$}
\newcommand{\het}{\it HET~}

\begin{document}
\bibliographystyle{/Active/my2}

\title{The Mass of \HD c from {\it Hubble Space Telescope} Astrometry and High-Precision Radial Velocities\footnote{Based on observations made with the NASA/ESA Hubble Space Telescope, obtained at the Space Telescope Science Institute, which is operated by the Association of Universities for Research in Astronomy, Inc., under NASA contract NAS5-26555. Based on observations obtained with the Hobby-Eberly Telescope, which is a joint project of the University of Texas at Austin, the Pennsylvania State University, Stanford University, Ludwig-Maximilians-UniversitŠt MŸnchen, and Georg-August-Universit\"{a}t G\"{o}ttingen.} }

\author{ G.\ Fritz Benedict\altaffilmark{2}, Barbara E.
McArthur\altaffilmark{2}, Jacob L. Bean\altaffilmark{3}, Rory Barnes\altaffilmark{4}, Thomas E. Harrison\altaffilmark{5},  Artie 
Hatzes\altaffilmark{6}, Eder Martioli\altaffilmark{7}, and Edmund P. Nelan\altaffilmark{8}  }

\altaffiltext{2}{McDonald Observatory, University of Texas, Austin, TX 78712}
\altaffiltext{3}{Institut f\"{u}r Astrophysik, Georg-August-Universit\"{a}t G\"{o}ttingen, Friedrich-Hund-Platz 1, 37077 G\"{o}ttingen, Germany, Marie Curie International Incoming Fellow}
\altaffiltext{4}{University of Washington, Seattle, WA 98195}
\altaffiltext{5}{New Mexico State University, Las Cruces, NM 88003}
\altaffiltext{6}{Thuringer Landessternwarte, Tautenburg, D-07778 Germany}
\altaffiltext{7}{Instituto Nacional de Pesquisas Espaciais, S. J. dos Campos, SP Brazil}
\altaffiltext{8}{Space Telescope Science Institute, 3700 San Martin Dr., Baltimore, MD 21218}




\begin{abstract}
{\it Hubble Space Telescope} (\HST) Fine Guidance Sensor astrometric observations of the G4 IV star 
\HD ~are combined with the results of the analysis of extensive ground-based radial 
velocity  data to determine the mass of the outermost of two previously known companions.     
Our new radial velocities obtained with the Hobby-Eberly Telescope and velocities from the Carnegie-California group  now span over eleven years. With these data we obtain  improved RV orbital elements for both the inner companion, \HD b and the outer companion, \HD c.  
We identify a rotational period of \HD~(P$_{rot}=31.65 \pm 0.17^ d$) with FGS photometry. The inferred star spot  fraction is consistent with the remaining scatter in velocities being caused by spot-related stellar activity. We then model the combined astrometric  and RV measurements
 to obtain the parallax, proper 
motion, perturbation period, perturbation inclination, and perturbation size due to \HD c. For \HD c we find  
P = 2136.1 $\pm$ 0.3 d,  perturbation semi-major axis $\alpha =1.05 \pm 0.06$ mas, and  
inclination $i$ = 48.3\fdg0 $\pm$ 4\fdg0.   Assuming a primary mass 
$M_* = 1.48 M_{\sun}$, we obtain a companion mass ${\it M}_c = 17.6 ^{ +1.5}_{-1.2} {\it 
M}_{Jup}$, $3\sigma$  above a 13 \mjup deuterium burning, brown dwarf lower limit.   Dynamical simulations incorporating this accurate mass for \HD c indicate that a near-Saturn mass planet could exist between the two known companions.  
We find weak evidence of an additional low amplitude signal that can be modeled as a planetary-mass ($\sim$0.17{\it 
M}$_{Jup}$) companion at P$\sim194$ days.  Including this component in our modeling  lowers the error of the mass determined for \HD c. Additional observations (radial velocities and/or {\it Gaia} astrometry) are required to validate
an interpretation of \HD d as a planetary-mass companion. If confirmed, the resulting \HD~planetary system may be an
example of a ``Packed Planetary System".

\end{abstract}


\keywords{astrometry --- interferometry --- stars: individual (\HD) ---  stars: 
radial 
velocities --- stars: late-type --- stars: distances --- extrasolar planets: 
masses --- brown dwarfs}


%

\section{Introduction}

\HD (= HIP 27253 =  HR 1988 = PLX 1320) hosts  two known companions discovered by high-precision radial velocity (RV) monitoring \citep{Fis01,Fis03,Wri09}. Previously published periods were P$_b$=14.31$^d$ and P$_c = 2146^d$ with minimum masses $M_b sini = 0.85$\mjup and $M_c sini = 13.1$\mjupe, the latter right above the currently accepted brown dwarf mass limit.
A predicted minimum perturbation for the outermost companion, \HD c, $\alpha_c = 0.8$ millisecond of arc (mas), motivated us to obtain millisecond of arc per-observation precision astrometry with {\it HST}  with which to determine its true mass (not the minimum mass, $M_c sini$).  These astrometric data now span 3.25 years.  

In the early phases of our project \cite{Ref06} derived an estimate of the mass of \HD c from \HIP, obtaining  ${\it M}_c = 38 ^{ +36}_{-19} $ \mjupe, well within the brown dwarf 'desert'. Recent comparisons of FGS astrometry with \HIP, e.g. \cite{lee07b}, suggest that we should obtain a more precise and accurate mass for \HD c. Our mass is derived from 
combined astrometric and RV data, continuing a series 
presenting accurate masses of planetary, brown dwarf, and non-planetary companions to nearby 
stars. Previous results include the mass of Gl 876b \citep{Ben02c}, of 
$\rho^1$ Cancri d \citep{McA04}, $\epsilon$ Eri b \citep{Ben06}, HD 33636B \citep{Bea07}, and HD 136118 b \citep{Mar10}.
 
\HD~is a metal-rich G4 IV star at a distance of about 40 pc. The star lies in the 'Hertzsprung Gap' (Murray \& Chaboyer 2002), a region typically traversed very quickly as a star evolves from dwarf to giant. \cite{Bai08a} have measured a radius. \HD~also has a small IR excess found by \citet{Mor07} with {\it Spitzer} and interpreted as a Kuiper Belt at 20--50 AU from the primary.
Stellar parameters are summarized in Table~\ref{tbl-STAR}. 

In Section 2 we model  RV data from four sources, obtaining orbital parameters for both \HD b and \HD c. We also discuss and identify RV noise sources.  In Section 3 we present the results of our combined astrometry/RV modeling, concentrating on \HD c.  We briefly discuss  the quality of our astrometric results as determined by residuals, and derive an absolute parallax and relative proper motion for \HD, those nuisance parameters that must be removed to determine the perturbation parameters for the perturbation due to component c.  Simultaneously we derive the astrometric orbital parameters. These, combined with an estimate of the mass of \HD, provide a mass for \HD c. Section 4 contains the results of searches for additional components, limiting the possible masses and periods of such companions. In Section 5 we discuss possible identification of an RV signal that remained after modeling components b and c. We discuss our results and summarize our conclusions in Section 6.

\section{Radial Velocities}
 \subsection{RV Orbits} \label{RVsec}
 We first model RV data, a significant fraction of which comes from the Hobby-Eberly Telescope (HET).  Measurements  from the California-Carnegie exoplanet research group \citep{Wri09} and a few from the McDonald Harlan J Smith telescope \citep{Wit09b} were also included. The California-Carnegie data were particularly valuable,  increasing the time span from four to over eleven years. Our astrometry covers a little only 70\% of the orbit of \HD c, and in the absence of a multi-period span of radial velocities, would not be sufficient to establish accurate perturbation elements, particularly period, eccentricity, and periastron passage. All RV sources are listed in Table~\ref{tbl-RV}, along with the RMS of the residuals to the combined orbital fits described below. The errors for all published RV and our new HET RV have been modified by adding in quadrature the expected RV jitter from stellar activity determined in Section~\ref{SR}. 
 
 All the radial velocities were obtained using I$_2$ cell techniques. The HET  data were obtained with the HET High-resolution Spectrograph (HRS), described in \cite{Tul98} and processed with the I$_2$ pipeline described in \cite{Bea07}, utilizing robust estimation to combine the all velocities from the individual chunks. Typically three HET observations are secured within 10--15 minutes. These are combined using robust estimation to form normal points for each night. The HET normal points and associated 
 errors are listed in Table~\ref{tbl-HETRV}. 


 Combining RV observations from different sources is possible in the modeling environment we use. GaussFit \cite{Jef88} has the capability to simultaneously solve for many separate velocity offsets (because velocities from different sources are relative, having differing  zero points), along with the other orbital parameters. Relative  offsets ($\gamma$) and associated errors are listed in Table~\ref{tab:allorb3}. 

\begin{figure}
\epsscale{1.00}
\plotone{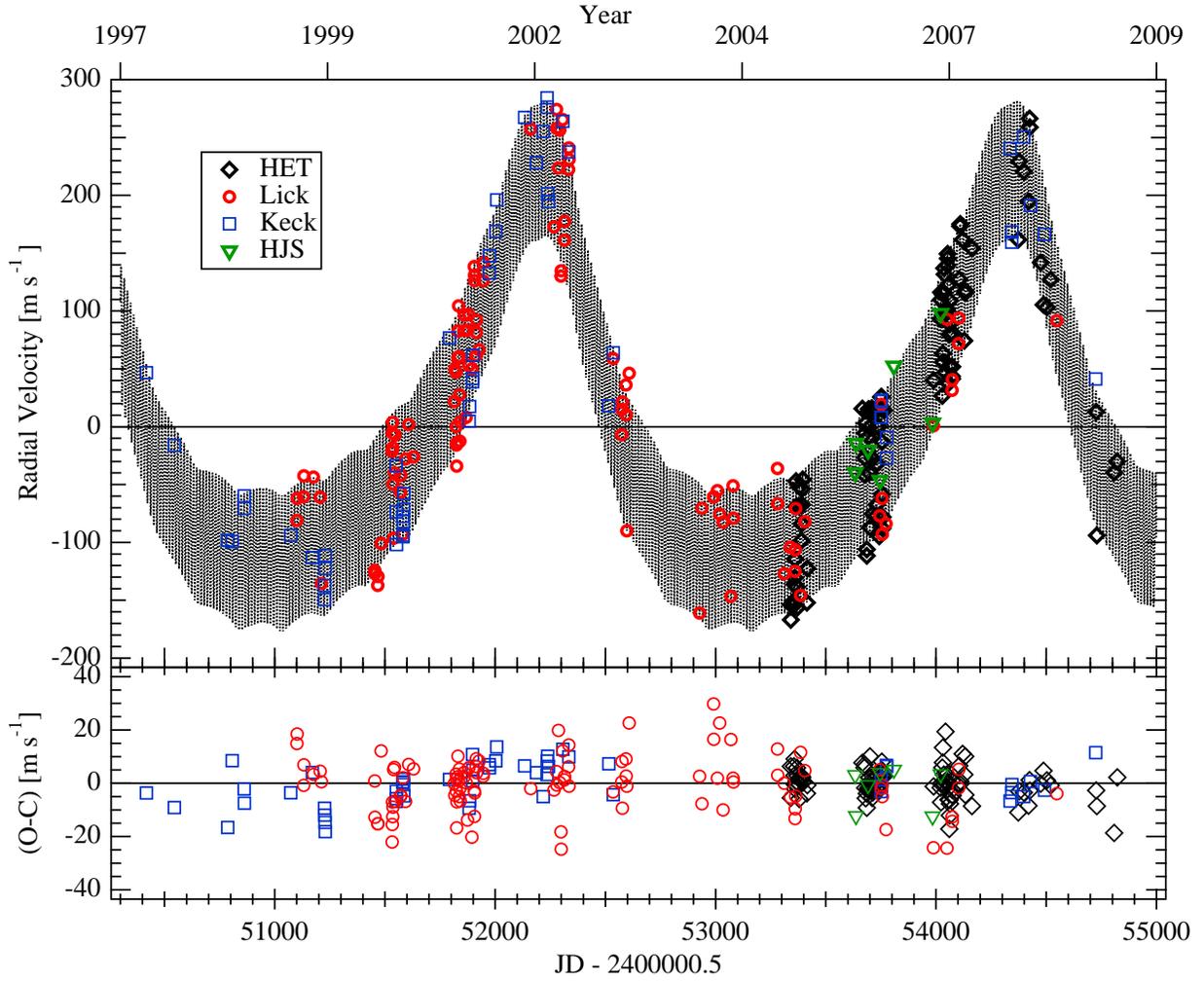}
\caption{RV measurements of \HD ~from sources as indicated in the legend 
(and identified in Table~\ref{tbl-RV}).  The  line is the velocity predicted from the orbital parameters (Table~\ref{tab:allorb3}) 
derived in the combined solution. Residuals (RV observed  minus RV calculated from the orbit) are plotted at bottom.} \label{fig-RVSa}
\end{figure}

\begin{figure}
\epsscale{0.85}
\plotone{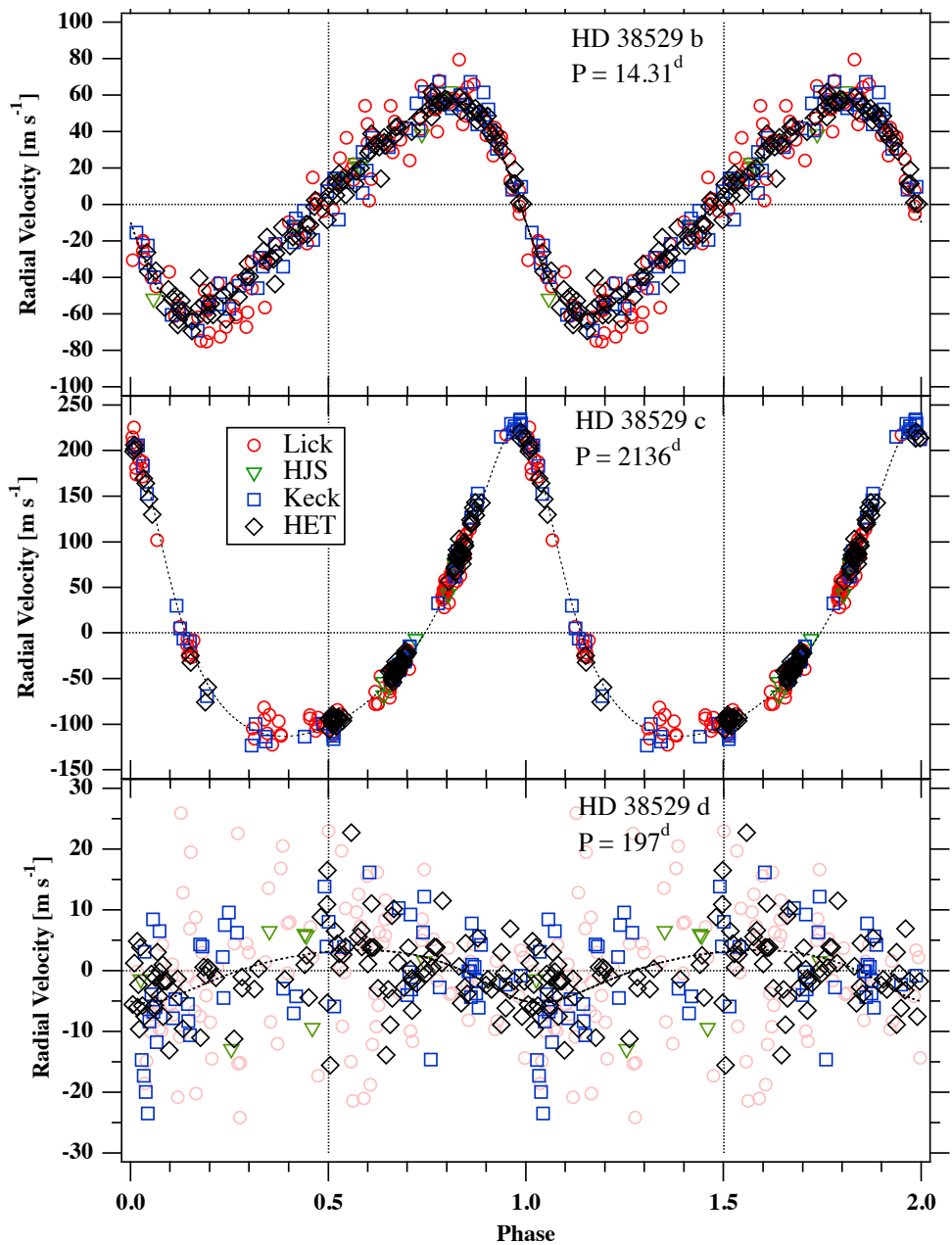}
\caption{RV measurements of \HD ~from sources as indicated in the legend 
(and identified in Table~\ref{tbl-RV}) phased to the orbital periods determined from a 
combined solution including astrometry and RV (Section~\ref{combo}).  
The dashed line is the velocity predicted from the orbital parameters (Table~\ref{tab:allorb3}) 
derived in the combined solution.} \label{fig-RVS}
\end{figure}

Orbital parameters derived from a combination of HET, HJS, Lick, and Keck radial velocities and \HST~astrometry will be provided in Section~\ref{combo}. Figure~\ref{fig-RVSa} shows the entire span of data along with the best fit multiple-Keplerian orbit. We note that there is sufficient bowing in the residuals to justify continued low-cadence RV monitoring, particularly given the prediction of \cite{Mor07} of dynamical stability for planets with periods as short as $\sim 70$y. Subtracting in turn the signature of first one, then the other known companion from the original velocity data we obtain the component b and c RV orbits shown in Figure~\ref{fig-RVS}, each phased to the relevant periods.   Re-iterating, all RV fits were modeled simultaneously with the astrometry.

Compared to the typical perturbation RV curve (e.g. Hatzes \etal 2005, McArthur \etal 2004, Cochran \etal 2004)\nocite{Hat05}\nocite{McA04}\nocite{Coc04}, our original orbits for components b and c exhibited significant scatter, much due to the identified stellar noise source discussed in Section~\ref{SR} below. Periodogram analysis of RV residuals to simultaneous fits of components b and c indicated a significant peak with a period near 197 days. The existence of the signal is fairly secure. A bootstrap analysis carried out by randomly shuffling the RV residual values 200,000 times
(keeping the times fixed), and determining if the random data
periodogram had peaks higher than the real data periodogram in the
frequency range $0 < \nu < 0.02 ~{\rm day}^{-1}$, yielded a false alarm probability, FAP =$5\times 10^{-4}$.  This motivated the addition of a third Keplerian component, resulting in the fit shown in the bottom panel of Figure~\ref{fig-RVS}. Even though the amplitude of this signal is about that expected from stellar noise, including this component (five additional parameters) in the combined modeling improved both the reduced $\chi^2$, and the RMS scatter as shown in Table~\ref{tbl-RV}. Identification of the cause of the signal will be discussed further in Section~\ref{dornod}.

\subsection{Stellar Rotation and the RV Noise Level} \label{SR}
 There are a number of  sources of RV noise intrinsic to \HD:  pulsations and velocity perturbations introduced by star spots and/or plages. The velocity effects caused by the latter two are modulated by stellar rotation. \cite{Val05} measure a rotation of \HD, V$_{rot}\,sin\,i =3.5 \pm 0.5$\kms.  \HD~is subgiant star, evolving towards the giant branch of the Hertszprung-Russell diagram, and is expected to have a higher level of pulsational activity than a main sequence star \citep{Hat08}. The pulsational amplitude can be estimated using the scaling relationship of \cite{Kje95}
$V_{amp} = (L/L_{\sun})/(M/M_{\sun}) \times 0.234$ \ms.
The luminosity and mass of \HD~yield  a pulsational amplitude of $\sim1$ \ms, so this alone cannot
account for the excess RV scatter. 

\begin{figure}
\epsscale{1.00}
\plotone{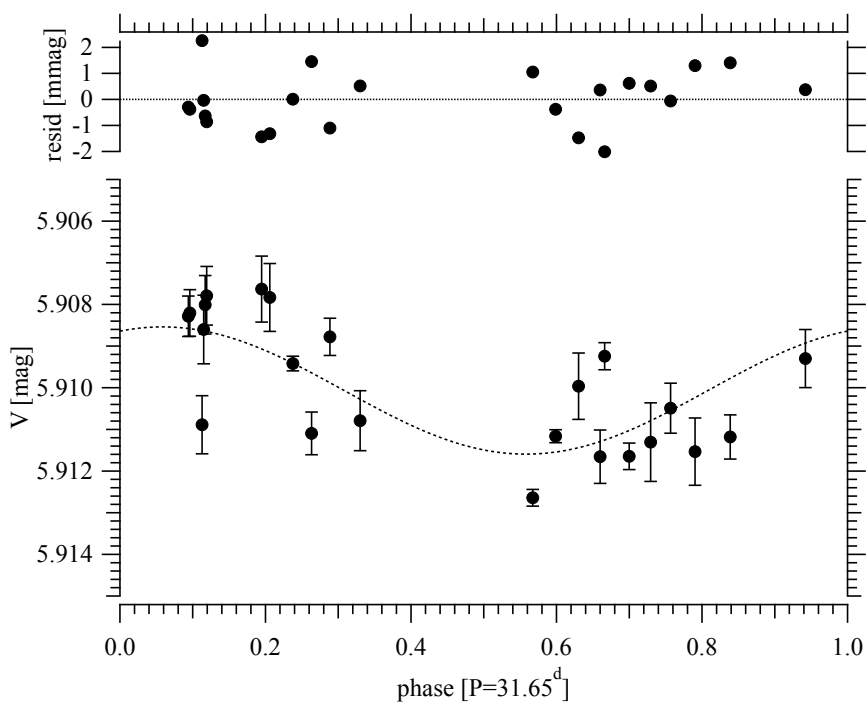}
\caption{FGS photometry of \HD~phased to a period, P=31.65$^d$. \HD~intensities were flat-fielded with average counts from all astrometric reference stars observed during each observation set. Normal points are formed from the five observations of \HD~secured within each set. The zero-point is chosen so that the average V magnitude matches the SIMBAD value. The amplitude of the photometric variation is 0.15\%. A periodogram of the flat field values shows no significant signals in the range $10 < P < 70^d$.} \label{fig-phot}
\end{figure}

Several relationships between the amplitude of RV noise and the fraction of star spot coverage have been developed. \cite{Sar97} obtain $A_{RV} = 6.5\times v\,sin\,i \times f^{0.9}$, where $f$ is the spot filling factor in percent. \cite{Hat02} obtained $A_{RV} = (8.6v\,sin\,i - 1.6) \times f^{0.9}$. We can estimate the spot filling factor from FGS photometry of \HD. The FGS has been shown to be a photometer precise at the 2 millimag (mmag) level \citep{Ben98a}. We flat-fielded the \HD~FGS photometry, using an average of the counts from the astrometric reference stars listed in Table~\ref{tbl-1}, and plotted it against time. Clearly not constant at a level ten times our internal precision, a Lomb-Scargle periodogram showed a significant period at P=31.6 days (FAP=$4.3\times 10^{-4}$). A sin wave fit to the photometry yielded
P=$31.65 \pm 0.17^ d$ with an amplitude = 1.5 $\pm$ 0.2 mmag. Figure~\ref{fig-phot} is a plot of these photometric data phased to that period. 

The \cite{Val05} V$_{rot}\,sin\,i =3.5$\kms and a stellar radius from \cite{Bai08a}, R=2.44$ \pm $0.22 R$_{\sun}$, would predict a minimum P$_{rot}=32\pm 5^d$. Interpreting the modulation period of 31.6 days as the stellar rotation period, we ascribe the photmetric variation (0.15\%) to rotational modulation of star spots. The photometric amplitude suggests an RV noise level of 4--5 \ms. Taking the HET velocity RMS as closer to the true RV variation, we identify the remaining RV scatter as a combination of the three effects identified.


\section{\HST  Astrometry}

We used {\it HST} Fine Guidance Sensor 1r (FGS1r) to carry out our space-based  
astrometric observations. Nelan \etal (2007)\nocite{Nel07}  provides a detailed 
overview of FGS1r as a science instrument.  \cite{Ben02b, Ben06} 
describe the FGS3 instrument's astrometric capabilities along with the data acquisition and 
reduction strategies used in the present study. We use FGS1r for the present study because 
it provides superior fringes from which to obtain target and reference star 
positions  \citep{McA02}.

\begin{center}
\begin{figure}
\epsscale{0.75}
\plotone{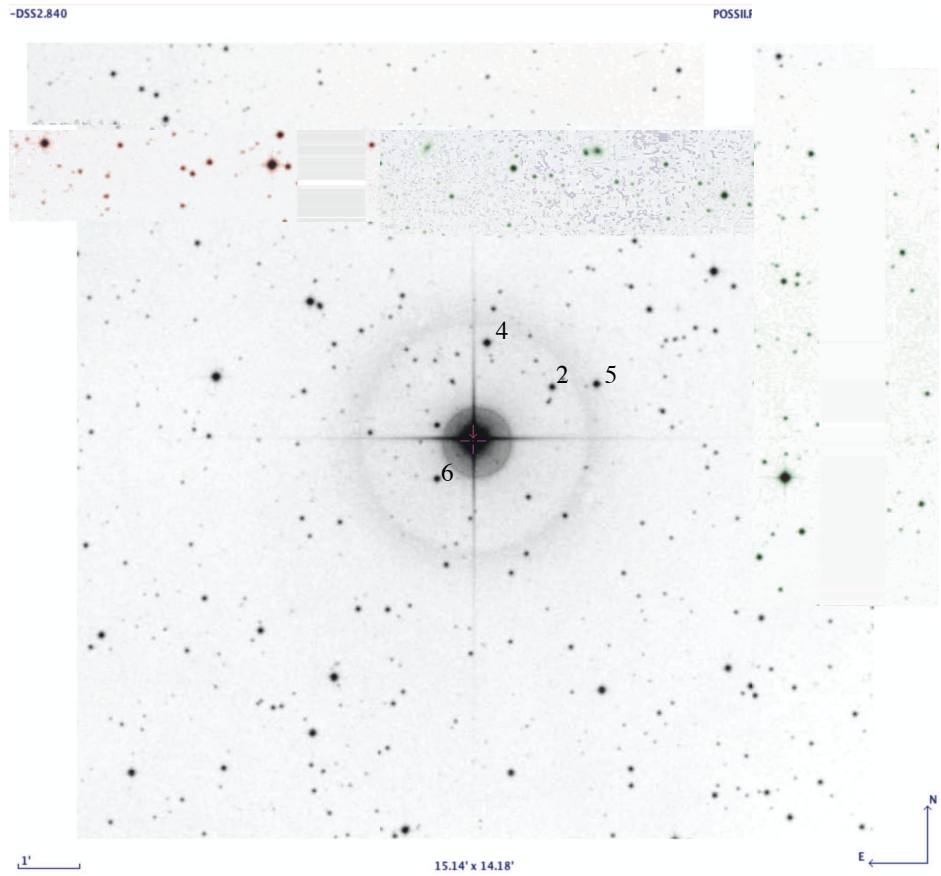}
\caption{Positions of \HD~(center) and the astrometric reference stars identified in Table~\ref{tbl-1}.}
\label{fig-Find}
\end{figure}
\end{center}

\begin{center}
\begin{figure}
\epsscale{1.00}
\plotone{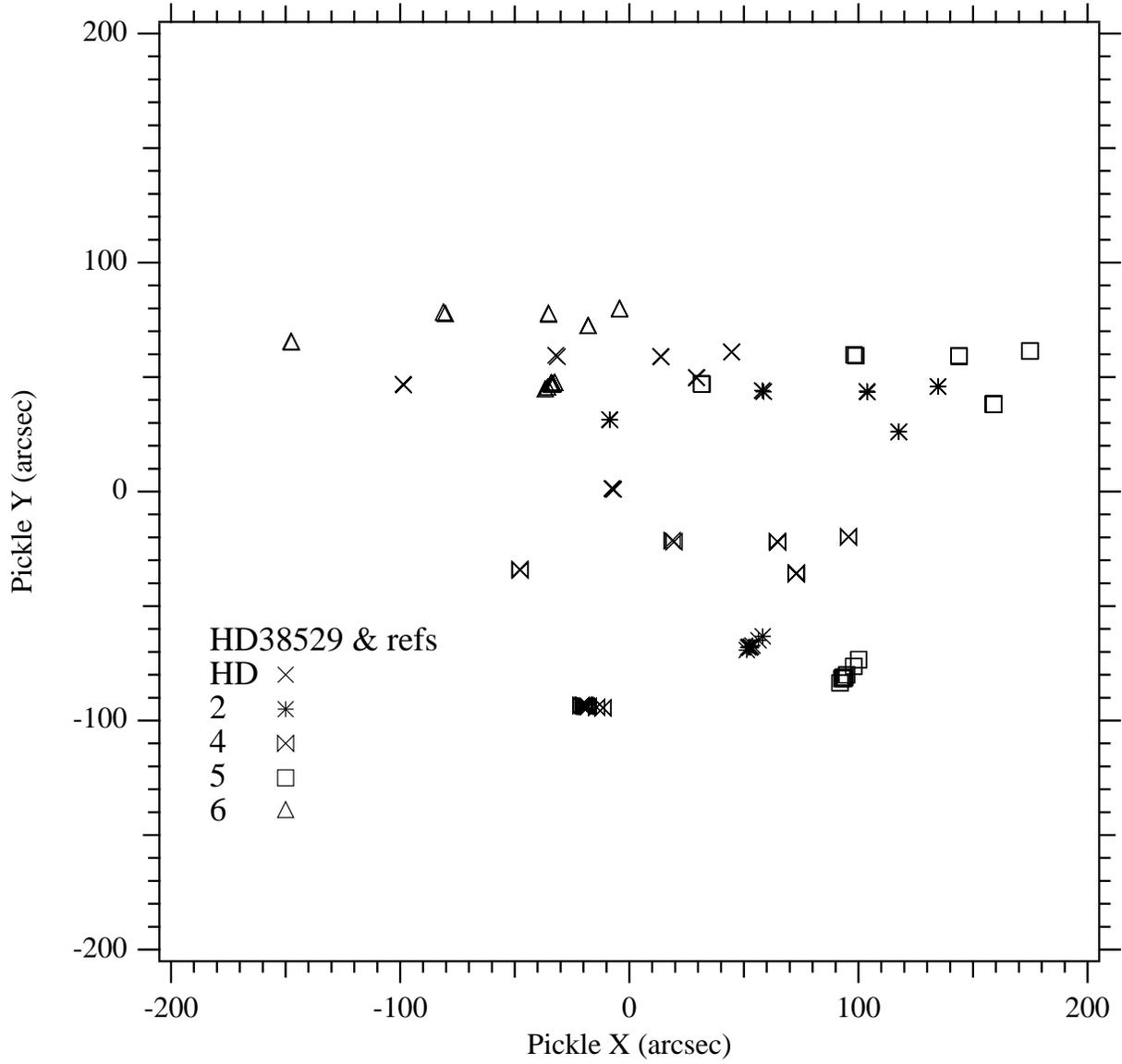}
\caption{Positions of \HD~and astrometric reference stars in FGS1r `pickle' coordinates. Due to two-gyro guiding constraints and guide star availability it was not possible to keep \HD~in the pickle center at each epoch.  }
\label{fig-Pick}
\end{figure}
\end{center}

\HD~is shown in Figure~\ref{fig-Find} along with the astrometric reference stars used in this study. Table~\ref{tbl-LOOF} presents a log of \HST FGS observations.  Note the bunching of the observation sets, each `bunch' with a time span less than a few days. Each set is tagged with the time of the first observation within each set. 
The field was observed at a very limited range of spacecraft roll 
values. As shown in Figure~\ref{fig-Pick}  \HD ~had to be placed in different locations within the FGS1r FOV 
to accommodate the distribution of astrometric reference stars and to insure availability of guide stars required by the other two FGS units. Additionally,  all observation sets suffered from observation timing constraints imposed by two-gyro guiding\footnote{\HST has a full compliment of six rate gyros, two per axis, that provide coarse pointing control. By the time these observations were in progress, three of the gyros had failed. \HST can point with only two. To ``bank" a gyro in anticipation of a future failure, NASA decided to go to two gyro pointing as standard operating procedure.}. 
Note that due to the extreme bunching of the epochs, we acquired effectively only five astrometric epochs. Also, we note that the last group of observation sets were a `bonus'. In November 2008 the only science instrument operating on \HST was FGS1r. Consequently, we were able to acquire additional observation sets for a few of our prime science targets, including \HD.  These recent data significantly lengthened the time span of our observations, hence, increased the precision with which the parallax and proper motion could be removed to determine the perturbation orbit of \HD.  Once combined with an estimate of the mass of \HD, the perturbation size will provide  the mass of the companion, \HD c.

\subsection{\HD~Astrometric Reference Frame}  \label{AstRefs}
The astrometric reference frame for \HD~consists of four stars. Any prior knowledge concerning these four stars  
eventually enters our modeling as observations with error, and yields the most 
accurate parallax and proper motion for the prime target, \HD. These periodic and non-
periodic motions must be removed as accurately and precisely as possible to obtain the 
perturbation inclination and size caused by \HD c. Of particular value are independently measured proper motions. This particular prior knowledge comes from the UCAC3 catalog \cite{Zac09}.
Figure \ref{fig-Pick} shows the distribution in FGS1r pickle coordinates of the 23
 sets of four reference star measurements for the \HD ~field.  At each epoch we measured each reference stars 2 -- 4 
times, and \HD~five times.

\subsection{Absolute Parallaxes for the Reference Stars}
Because the parallax determined for \HD~is measured with respect to 
reference frame stars which have their own parallaxes, we must either apply a statistically-
derived correction from relative to absolute parallax \citep[Yale Parallax Catalog, YPC95]{WvA95}, adopt an independently derived parallax (e.g., {\it Hipparcos}), or 
estimate the absolute parallaxes of the reference frame stars.  In principle, 
the colors, spectral type, and luminosity class of a star can be used to estimate the 
absolute magnitude, $M_V$, and $V$-band absorption, $A_V$. The absolute parallax for each reference star is then 
simply,
\begin{equation}
\pi_{\rm abs} = 10^{-(V-M_V+5-A_V)/5}
\end{equation}
\subsubsection{ Reference Star Photometry and Spectroscopy}
Our band passes for reference star photometry include: $BVRI$ photometry of the reference stars from the NMSU 1 m
telescope located at Apache Point Observatory and JHK (from 2MASS\footnote{The Two Micron All Sky Survey
is a joint project of the University of Massachusetts and the Infrared Processing
and Analysis Center/California Institute of Technology }). 
Table \ref{tbl-SPP} lists the visible and infrared photometry for the \HD  ~reference stars.
The spectra from which we estimated spectral type and luminosity class were obtained on 2009 December 9 using the
RCSPEC on the Blanco 4 m telescope at CTIO. We used the KPGL1 grating
to give a dispersion of 0.95 \AA /pix. Classifications used a combination of template matching and line ratios.  The spectral types for the higher S/N stars are within $\pm$1 subclass. Classifications for the lower S/N stars are $\pm$2
subclasses. Table \ref{tbl-SPP} lists the spectral types and luminosity classes for our reference stars. 

\begin{center}
\begin{figure}
\epsscale{1.00}
\plotone{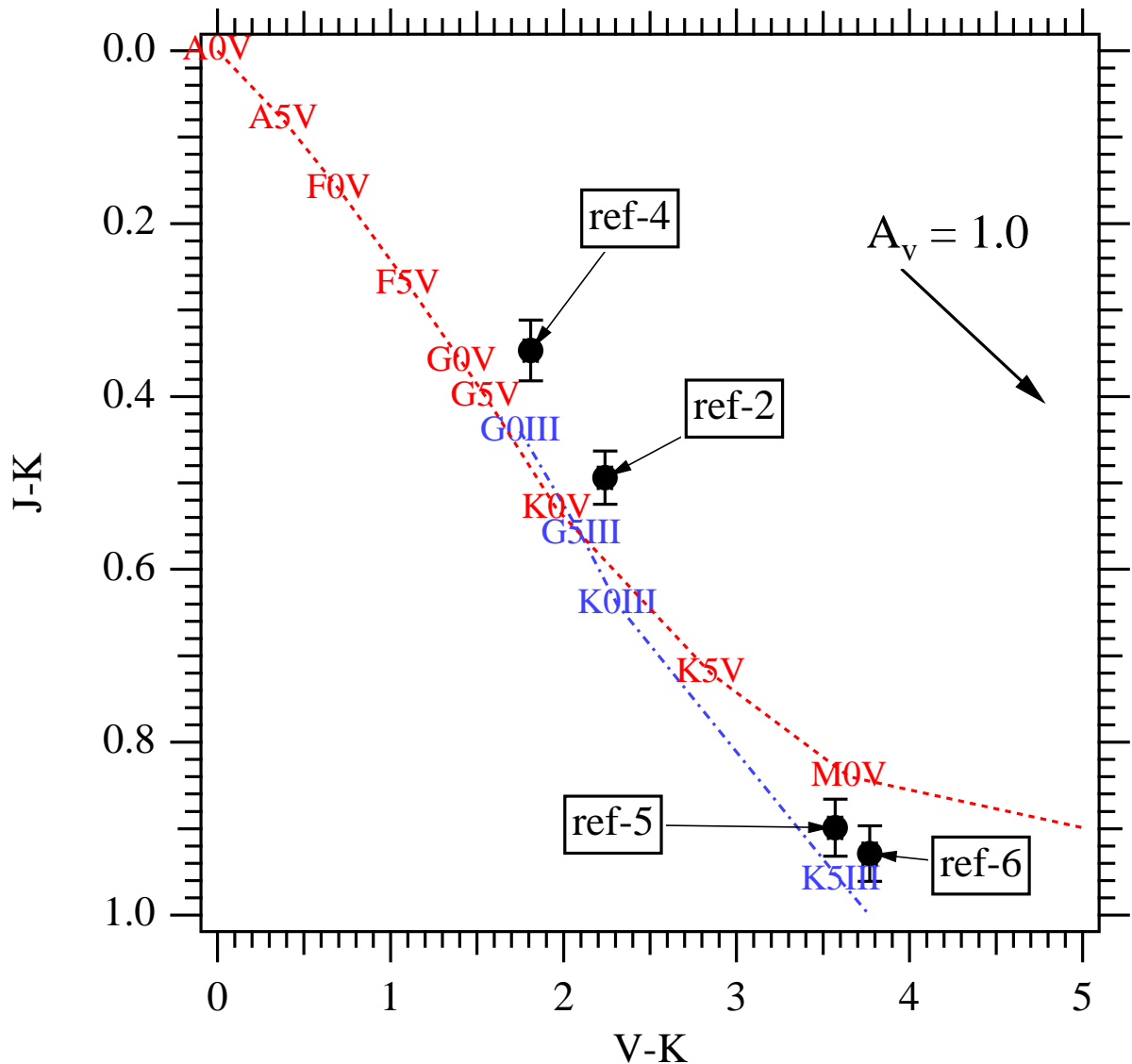}
\caption{$(J-K)$ vs. $(V-K)$ color-color diagram for stars identified in 
Table~\ref{tbl-1} and Table~\ref{tbl-IR}. The dashed line is the locus of  dwarf
(luminosity class V) stars of various spectral types; the dot-dashed line is for 
giants (luminosity class III). The reddening vector indicates $A_V = 1.0$ for the 
plotted color systems. Along this line of sight maximum extinction is  $A_V$$\sim$ 2 
\citep{Sch98}. }
\label{fig-CCD}
\end{figure}
\end{center}

Figure 
\ref{fig-CCD} contains a $(J-K)$ vs. $(V-K)$ color-color diagram of the  reference stars. Schlegel \etal (1998)  find an upper limit $A_V$$\sim$2 towards 
\HD, consistent with the absorptions we infer comparing spectra and photometry (Table~\ref{tbl-SPP}).

\begin{center}
\begin{figure}
\epsscale{0.75}
\plotone{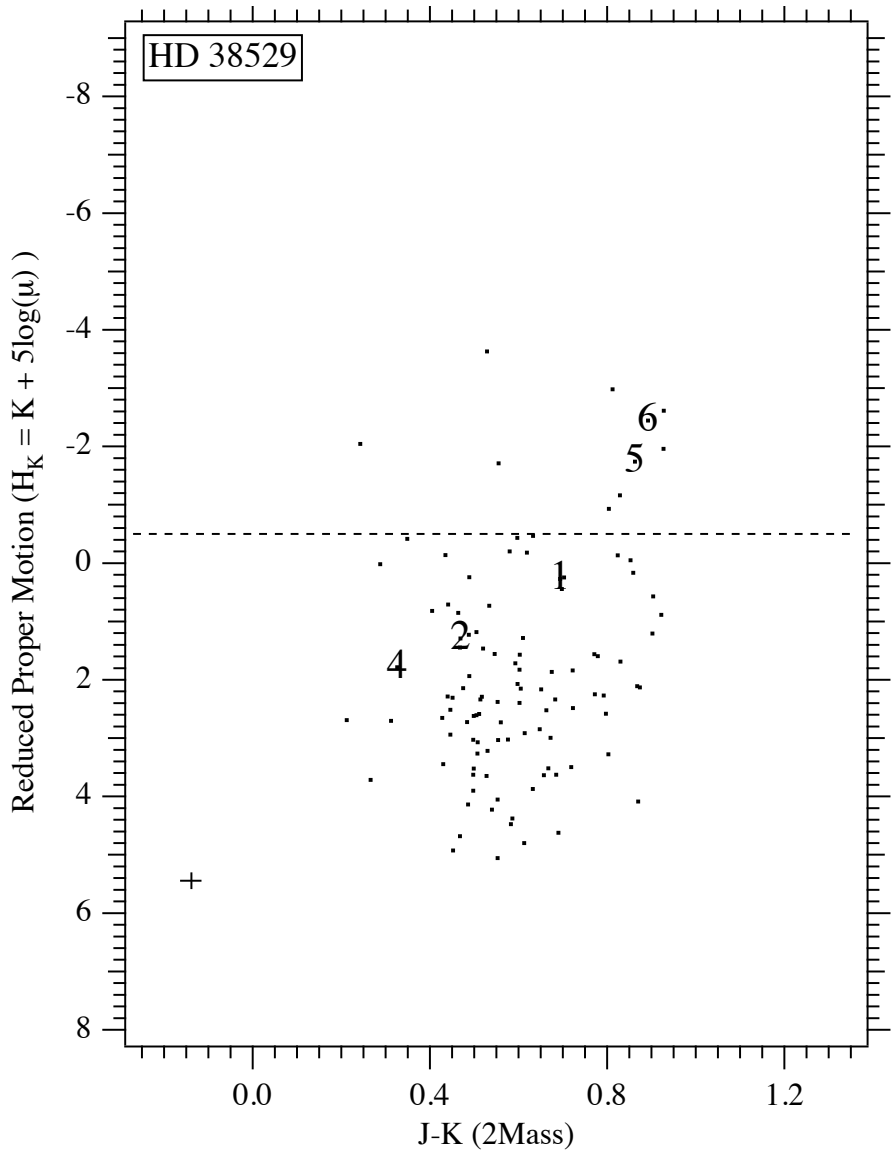}
\caption{Reduced proper motion diagram for 107 stars in a 1/3\arcdeg ~field centered
on \HD. Star identifications are in Table~\ref{tbl-1}. For a given spectral type, 
giants and sub-giants have more negative $H_K$ values and are redder than dwarfs in 
$(J-K)$.  $H_K$ values are derived from  proper motions in Table \ref{tbl-PM}. The small cross at the lower left represents a typical $(J-K)$ error of 0.04 mag and $H_K$ error of 0.17 mag.  Ref-5 and -6 are confirmed to have luminosity class III. \HD~(\#1 in plot) is also intermediate (luminosity class IV) in this parameter space.} 
\label{fig-RPM}
\end{figure}
\end{center}

The derived absolute magnitudes are critically dependent on the assumed stellar 
luminosity, a parameter impossible to obtain for all but the latest type stars 
using only  Figure \ref{fig-CCD}. To confirm the luminosity classes obtained from classification spectra we abstract  UCAC3 proper 
motions \cite{Zac09} for a one-degree-square field centered on \HD , and then 
iteratively employ the technique of reduced proper motion \citep{Yon03,Gou03}  to discriminate 
between giants and dwarfs. The end result of this process is contained in 
Figure~\ref{fig-RPM}. Reference stars ref-5 and ref-6 are confirmed to have luminosity class III (giant).

\subsubsection{Adopted Reference Frame Absolute Parallaxes}\label{CORR}

We derive absolute parallaxes by comparing our estimated spectral types and luminosity 
class to  $M_V$ values from \nocite{Cox00} Cox (2000). Our adopted input errors for 
distance moduli, $(m-M)_0$, are 0.5 mag for all reference stars. Contributions to the error are uncertainties in $A_V$ and errors 
in $M_V$ due to uncertainties in color to spectral type mapping.  All reference star absolute parallax estimates are listed in Table \ref{tbl-SPP}.  Individually, no reference star absolute parallax is better determined than ${\sigma_{\pi}\over \pi}$ = 
23\%. The average input absolute parallax for the reference frame is 
$\langle\pi_{abs}\rangle = 0.8$ mas. We compare this to the correction to absolute parallax discussed and presented in 
YPC95 (sec. 3.2, fig. 2).  Entering YPC95, fig. 2, with the Galactic latitude of \HD, $b = 
-19\arcdeg$, and average magnitude for the reference frame, $\langle V_{\rm ref} \rangle 
= 13.85$, we obtain a correction to absolute of 1.2 mas, consistent with our derived correction. As always \citep{Ben02a, Ben02b, Ben02c, McA04, Sod05, Ben06, Ben07}, rather than 
apply a model-dependent correction to absolute parallax, we introduce our 
spectrophotometrically-estimated reference star parallaxes  into our reduction 
model as observations with error.

\subsection{The Astrometric Model}
The \HD ~reference frame contains only four stars. From  
positional measurements we determine the scale, rotation, and offset ``plate
constants" relative to an arbitrarily adopted constraint epoch for
each observation set. As for all our previous astrometric analyses, we employ GaussFit (Jefferys \etal 
1988) 
\nocite{Jef88} to minimize $\chi^2$. The solved equations of condition for the 
\HD ~
field are:
\begin{equation}
        x^\prime = x + lc_x(\it B-V)  - \Delta XF_x
\end{equation}
\begin{equation}
        y^\prime = y + lc_y(\it B-V) - \Delta XF_y
\end{equation}
\begin{equation}
\xi = Ax^\prime + By' + C  - \mu_\alpha \Delta t  - P_\alpha\pi- ORBIT_x
\end{equation}
\begin{equation}
\eta = Dx^\prime + Ey^\prime + F  - \mu_\delta \Delta t  - P_\delta\pi- ORBIT_y
\end{equation}

\noindent for FGS1r data.
Identifying terms, $\it x$ and $\it y$ are the measured coordinates from {\it HST};   $(B-V)$ is the Johnson $(B-V)$ color of each star; and $\it lc_x$ and $\it lc_y$ are the lateral color corrections, applied only to FGS1r data. Here $\Delta XF_x$ and $\Delta XF_y$are cross filter corrections (see Benedict \etal 2002b) in $\it x$ and $\it y$, applied to the observations of \HD .  $A$, $B$, $D$ and $E$ are scale- and rotation plate constants, $C$ and $F$ are offsets; $\mu_\alpha$ and $\mu_\delta$ are proper motions; $\Delta t$ is the epoch difference from the constraint epoch; $P_\alpha$ and $P_\delta$ are parallax factors;  and $\it \pi$ is  the parallax.   We obtain the parallax factors from a JPL Earth orbit predictor 
(Standish 1990)\nocite{Sta90}, upgraded to version DE405. Orientation to the sky for the 
FGS1r data is obtained from ground-based astrometry (2MASS Catalog) with uncertainties 
of $0\fdg01$. $ORBIT_x$ and  $ORBIT_y$ are functions (through Thiele-Innes constants, e.g., Heintz, 1978) \nocite{Hei78} of the 
traditional astrometric and RV orbital elements listed in Table~\ref{tab:allorb3}.

\subsection{Assessing Reference Frame Residuals}
\begin{figure}
\epsscale{0.65}
\plotone{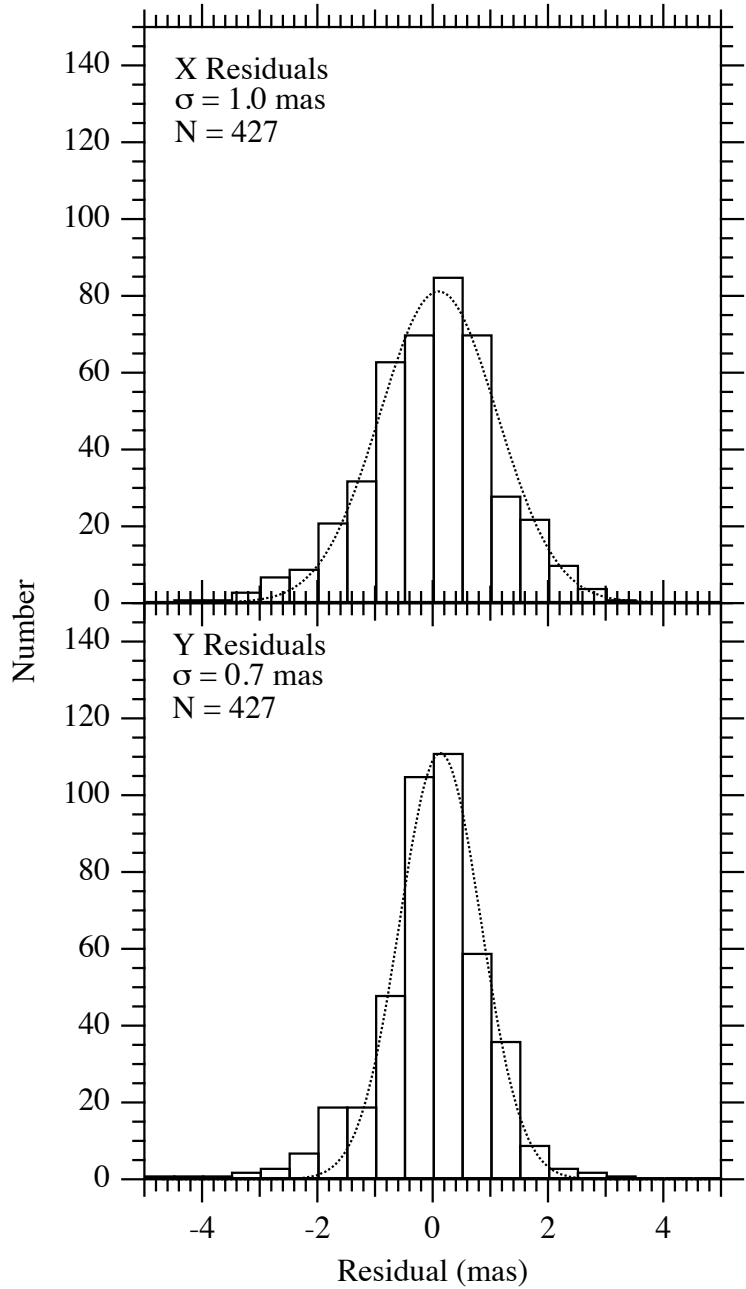}
\caption{Histograms of x and y residuals obtained from modeling the FGS 
observations of \HD ~and the FGS reference frame with equations 4 and 5. Distributions are 
fit with gaussians with standard deviations, $\sigma$, indicated in each panel.} \label{fig-FGSH}
\end{figure}

 The Optical Field Angle Distortion calibration \citep{McA02} 
reduces as-built {\it HST} telescope and FGS1r distortions with magnitude 
$\sim1\arcsec$ to below 2 mas  over much of the FGS1r field of regard. These data were calibrated with a revised OFAD generated by McArthur in 2007. From 
histograms of the FGS astrometric residuals (Figure~\ref{fig-FGSH}) we conclude 
that we have obtained correction at the $\sim 1$ mas level. The  reference frame 
'catalogs' from  FGS1r in $\xi$ and $\eta$ standard coordinates (Table \ref{tbl-POS}) 
were determined with  
$<\sigma_\xi>= 0.15$	 and	$<\sigma_\eta> = 0.15$ mas.

\subsection{The Combined Orbital Model} \label{combo}
We linearly combine unperturbed Keplerian orbit, simultaneously modelling the radial velocities and astrometry.  The period (P), the epoch of passage through periastron in years (T), the
eccentricity (e), and the angle in the plane of the true orbit between
the line of nodes and the major axis ($\omega$), are  the same for an orbit determined from
RV or from astrometry.  
The remaining orbital elements come only from astrometry. We force a relationship between the astrometry and the RV 
through this constraint \citep{Pou00}
\begin{equation}
\displaystyle{{\alpha~sin~i \over \pi_{abs}} = {P K (1 -
e^2)^{1/2}\over2\pi\times4.7405}} 
\end{equation}
\noindent where quantities derived only from astrometry (parallax, $\pi_{abs}$, 
primary perturbation orbit size, $\alpha$, and inclination, $i$) are on the left, and 
quantities  derivable from both (the period, $P$ and eccentricity, $e$), or radial 
velocities only (the RV amplitude of the primary, $K$), are on the right. In this case, given the fractional orbit coverage of the \HD c perturbation afforded by the astrometry,  all right hand side quantities are dominated by the radial velocities.


For the parameters critical in determining 
the mass of \HD ~we find a parallax, $\pi_{abs} = 25.11 \pm 0.19$ mas and a proper 
motion in RA of  $-78.60\pm 0.15$  mas y$^{-1}$ and in DEC of $-141.96 \pm 0.11$ mas y$^{-1}$. 
Table~\ref{tbl-SUM}  compares values for the parallax and proper motion of \HD 
~from {\it HST} and both the original \HIP~values and the recent  \HIP~re-reduction \citep{Lee07a}. We note satisfactory agreement for parallax and  proper motion. 
Our precision and extended study duration have significantly improved the accuracy and precision of the parallax and proper motion of \HD.

\begin{figure}
\epsscale{0.75}
\plotone{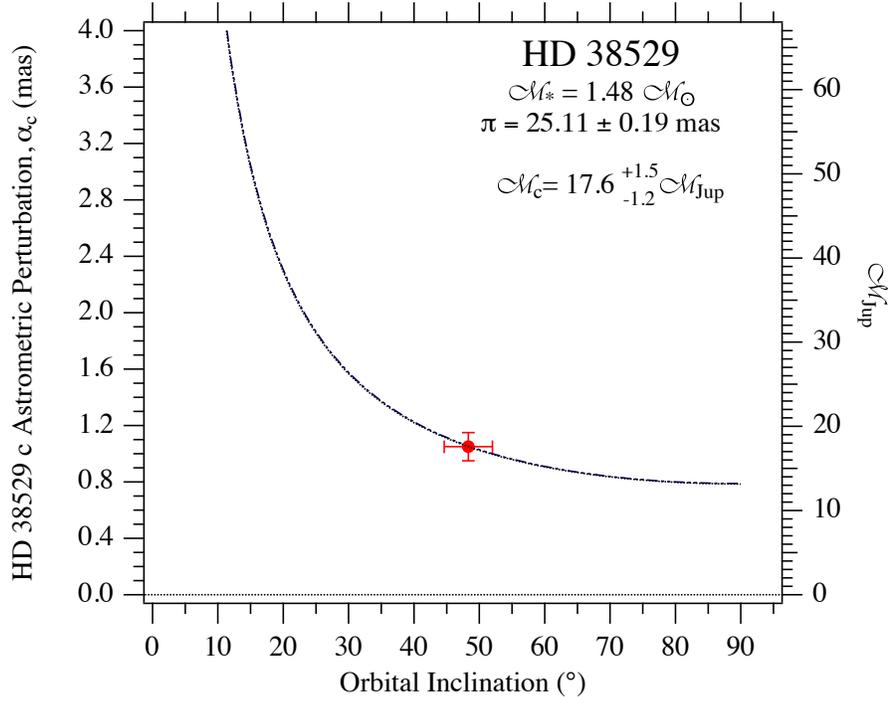}
\caption{This curve relates perturbation size and inclination for the \HD c perturbation through the Pourbaix \& Jorrisen (2000)  relation (Equation 6). We use the curve as a 'prior' in a quasi-bayesian sense. Our final values for the semimajor axis of the astrometric perturbation, $\alpha_c$, and inclination, $i_c$ are plotted with the formal errors.} \label{fig-PJconsTEMP}
\end{figure}

We find a perturbation size, $\alpha_c = 1.05 \pm 0.06$ mas, and an inclination, 
$i $= 48\fdg3 $\pm$ 3\fdg7. These, and the other orbital elements  are listed in Table~\ref{tab:allorb3} with 1-$\sigma$ errors. Figure~\ref{fig-PJconsTEMP} illustrates the Pourbaix and Jorrisen relation (Equation 6) between parameters 
obtained from astrometry (left-side) and radial velocities (right side) and our final  
estimates for $\alpha_c$ and $i$. 
In essence, our simultaneous solution uses the 
Figure~\ref{fig-PJconsTEMP} curve as a quasi-Bayesian prior, sliding along it until the 
astrometric and RV residuals are minimized. Gross deviations from the curve are 
minimized by the high precision of all of the right hand side terms in Equation 6 (Tables~\ref{tbl-SUM} and \ref{tab:allorb3}).   

\begin{figure}
\epsscale{0.75}
\plotone{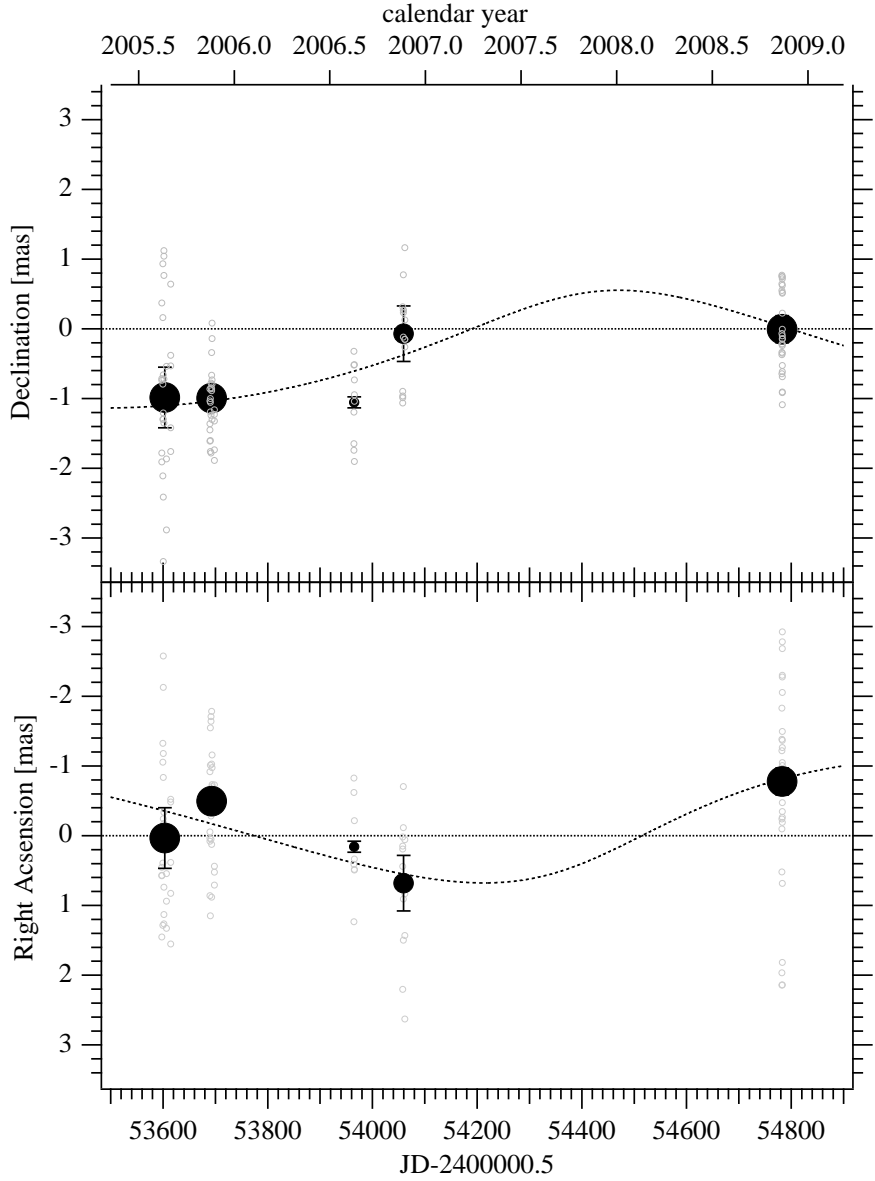}
\caption{RA (bottom) and Dec (top) components of the perturbation orbit for \HD c as a function of time.   The dashed line is the orbit described by the orbital elements found in Table~\ref{tab:allorb3}. The small symbols are all the \HD~observation residuals to a model that does not contain orbital motion.  The large symbols are normal points formed at each average epoch. Symbol size is proportional to the number of individual observations at each normal point epoch.} \label{fig-OrbXY}
\end{figure}
\begin{figure}
\epsscale{1.00}
\plotone{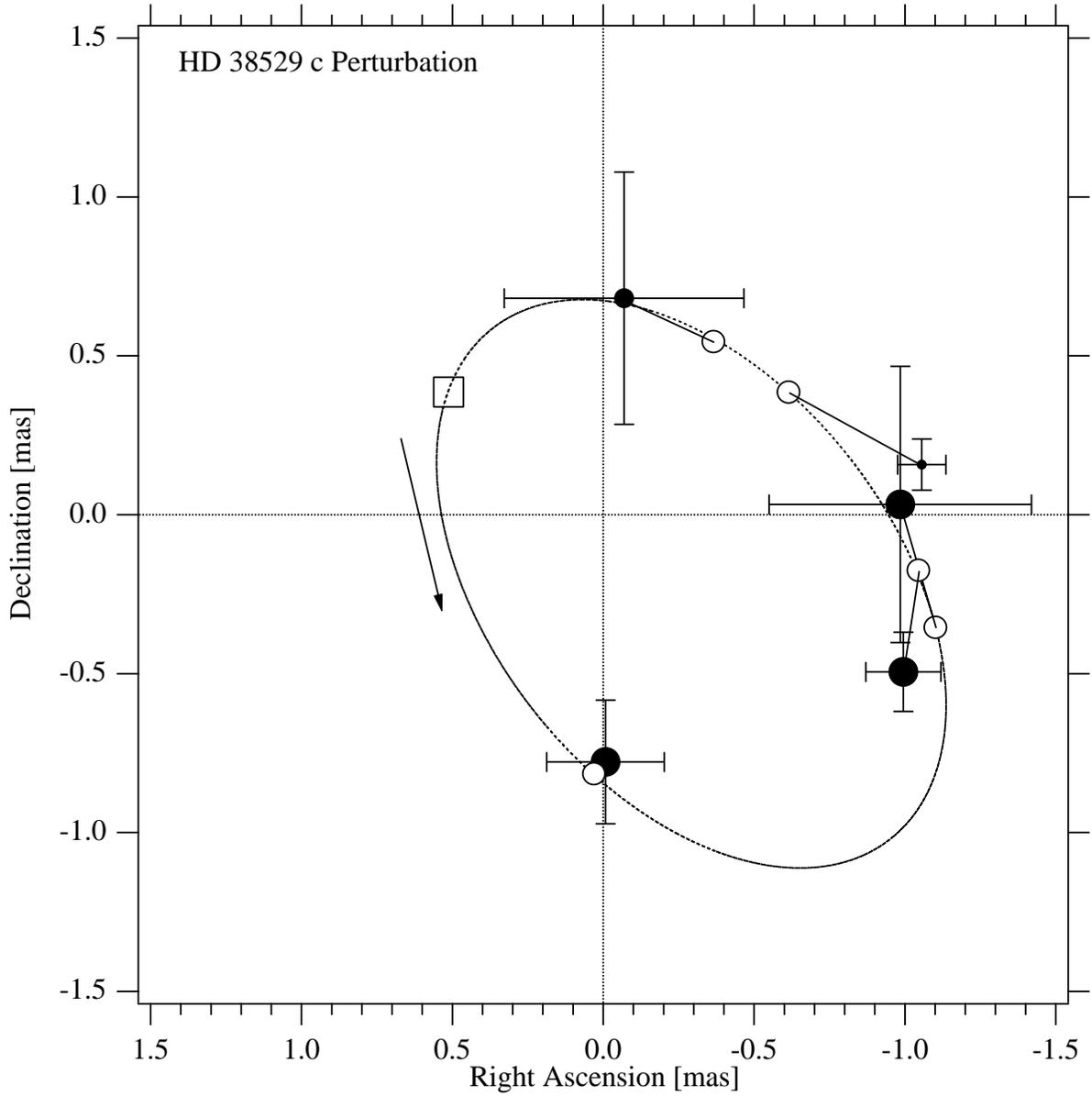}
\caption{Perturbation orbit for \HD c. Elements are found in Table~\ref{tab:allorb3}.  
Residual vectors are plotted, connecting each normal point residual to its predicted
position (O) at each epoch of observation. Error bars indicate the positional 1-$\sigma$ normal point dispersion within each epoch. Normal point symbol size is proportional to the number of individual observations at each normal point epoch. The open square marks periastron passage, T$_0$=2013.68, and the arrow indicates the direction of perturbation motion.} \label{fig-Orbc}

\end{figure}

At this stage we can assess the reality of the \HD c astrometric perturbation by plotting 
the astrometric residuals from a model that does not include a component c orbit. Figure~\ref{fig-OrbXY} shows the RA and Dec 
components of the  FGS residuals plotted as small symbols.  We also plot normal points formed from those smallest symbols within each of the 23 data sets listed in Table~\ref{tbl-LOOF}. The largest symbols denote the final normal points formed for each of our (effectively) five  epochs. Finally, each 
plot contains as a dashed line the RA and Dec components of the perturbation we find by 
including an orbit in our modeling.  Finally, Figure~\ref{fig-Orbc} shows the perturbation on the sky with our normal points. 

The planetary mass depends on the mass of the primary star, for which we have 
adopted $M_* = 1.48 M_{\sun}$ \citep{Tak07}. 
For that $M_*$ we find $M_c = 
17.6 ^{ +1.5}_{-1.2}$\mjupe, a significant improvement over the \cite{Ref06} estimate,  ${\it M}_c = 38 ^{ +36}_{-19} $ \mjupe, but agreeing within the errors. \HD c is likely a brown dwarf, but only about 3-$\sigma$ from the `traditional' planet-brown dwarf dividing line, 13\mjupe, the mass above which deuterium is thought to burn. In 
Table~\ref{tab:allorb3} the mass value, $M_c$, incorporates the present uncertainty 
in  $M_*$.  However, the dominant source of error is in the inclination estimate. Until \HD~c is directly detected, its radius is unknown. Comparing to the one known transiting brown dwarf, CoRot-Exo-3b \citep{Del08}, a radius of $R \sim 1 R_{JUP}$ seems reasonable.

\section{Limits on Additional Planets in the \HD~System} \label{lims}
The existence of  additional companions in the \HD~system is predicted by the ``Packed Planetary Systems" hypothesis \citep{Bar04,Ray05}. Specifically those investigations identified the range of orbits in which an additional planet in between planets b and c would be stable.
Having access to eleven years of \HD~RV observations permits a search for longer-period companions. Our velocity database, augmented by high-cadence ($\Delta$t  often less than 2 days, Table~\ref{tbl-HETRV}) HET monitoring, supports an exploration for shorter-period companions. Additionally,  a relatively precise actual mass for \HD c better informs any companion searches based on dynamical interaction. 


We independently examined the possible dynamical stability of an additional planet in the system by performing long-term N-body integrations of the orbits of the known planets and test particles in a manner similar to \citep{Bar04}. The orbital parameters of the known planets were taken to be those we have determined. Our advantage over previous stability investigations; the true mass of planet c was used. Planet b was assumed to be coplanar with planet c, and its mass was computed based on its minimum mass and the inclination of planet c. The test particles were initialized in orbits also coplanar with planet c, and with semimajor axes ranging from 0.01 to 10.0\,AU. The spacing was linear in the logarithm of the semimajor axis and 301 test particles were used. Simulations were done using three different eccentricity values for the test particles: e=0.0, 0.3. and 0.7. All the the calculations were carried out using the ``Hybrid'' integrator in the Mercury code \citep{Cha99}. The simulations were performed over 10$^{7}$\,yr and the integration parameters were tuned so that the fractional energy error was $<10^{-4}$.

\begin{center}
\begin{figure}
\epsscale{1.00}
\plotone{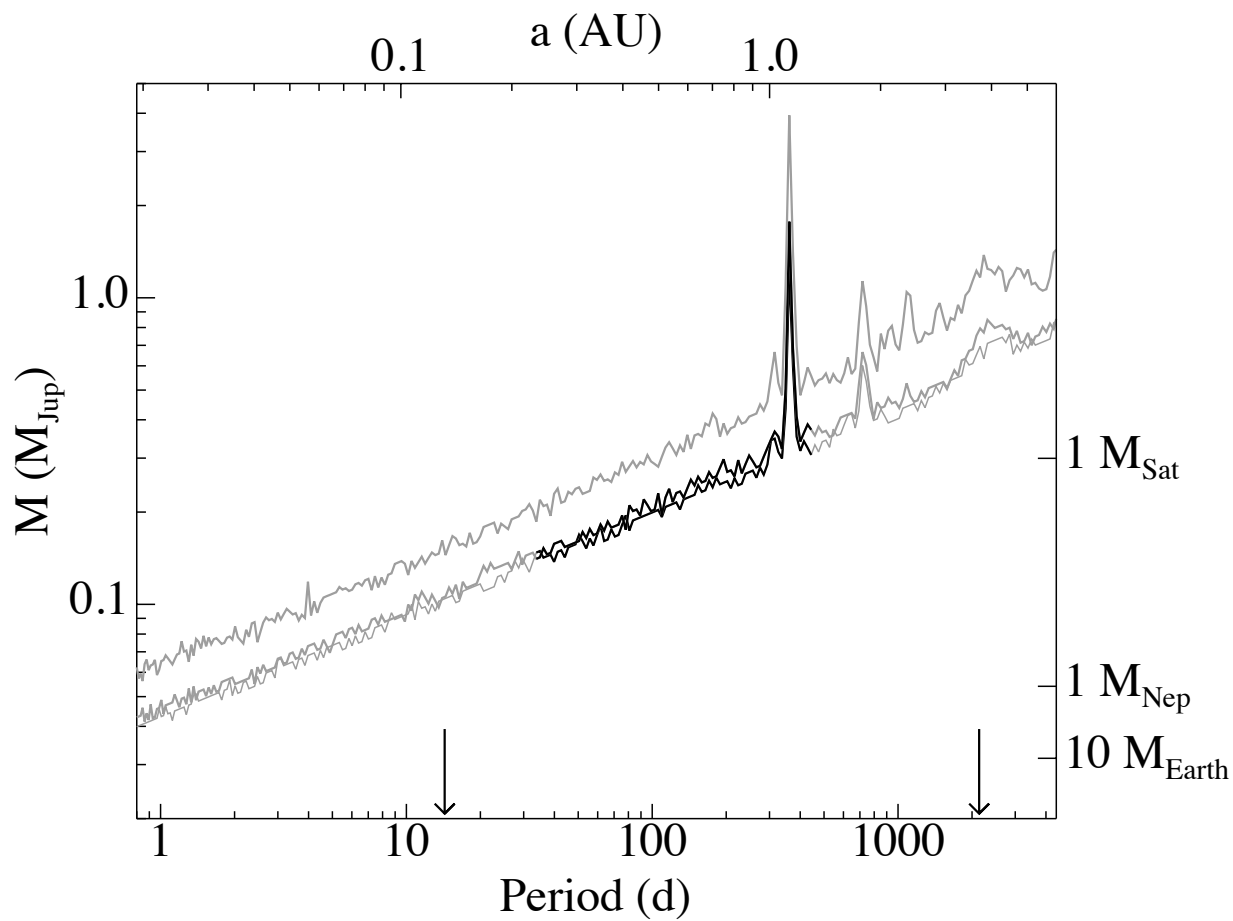}
\caption{Results from the detection limit simulations converted to mass of a companion assuming coplanarity with \HD c. Values above the lines would have been detected in a periodogram analysis of the RV data. The different lines represent the different assumed eccentricity values (lower: e=0.0, middle: e=0.3, upper: e=0.7). The black and grey parts of the lines indicate regions where test particles in the long-term simulation were stable and unstable respectively. The arrows indicate the orbital periods of the known planets.   }
\label{fig-JLBlim}
\end{figure}
\end{center}

From these simulations we find that no additional planets would be stable over long timescales interior to planet b. Between planets b and c, we find that planets with eccentricities less than 0.3 would be stable over the semimajor axis range 0.23 -- 1.32\,AU (P = 33 -- 455\,days). Exterior to planet c, no additional planets would be stable in orbits with periods shorter than the time baseline of the RV observations. Additional planets with eccentricities of 0.7 would not be stable over the entire range considered. These results are illustrated in Figure~\ref{fig-JLBlim}, and are completely consistent with the results of \cite{Bar04}. 

\section{Evidence for a Possible Tertiary, \HD d} \label{dornod}

We have previously mentioned (Section~\ref{RVsec}) a signal in the RV data residuals that remains after the signatures of components b and c are removed. An orbital fit to those residuals from the two component fit to \HD b and c
is shown in the bottom panel of Figure~\ref{fig-RVS}
with the  low-precision elements of this possible component d presented in Table~\ref{tab:allorb3}. 
Table~\ref{tab:allorb2} contains the orbital elements from a solution in which only components b and c are modeled.  While these astrometric elements closely agree with the three component solution in Table~\ref{tab:allorb3}, the errors are larger.
When \HD d is added
to the model (adding 5 degrees
of freedom, an increase of 1.5\%) the $\chi^2$ of the RV fit drops by 13\%, from 287 to 258. Comparing Tables~\ref{tab:allorb3} and \ref{tab:allorb2}, we see a similar reduction in the error in the mass of component c, the primary result of this study. \HD c remains a brown dwarf, whether or not component d is introduced.

We have explored sampling as a cause for the low-amplitude component d signal by performing power spectrum analysis of artificial RV generated from the
component b and c orbits in Table~\ref{tab:allorb2}, sampled on the dates of all the RV observations. There are no
peaks at P=194 days. To test for a seasonal effect in our HET data, one that might introduce a variation at the period of the tentative component d, we first removed the the large-amplitude component b and c signals. We then combined the residuals containing the low amplitude signal of \HD d with the  low amplitude signals found for HD 74156c \citep{Bea08a} and HD 136118c \citep{Mar10}.  The power spectrum of this aggregate should show seasonal fluctuations, if present. We saw nothing but expected signals (due to sampling) at 1/6, 1/2, and one year. 

Both the study in Section~\ref{lims} above and the packed planetary system hypothesis
\cite{Ray09}, allow this as a potential tertiary in the system. Furthermore, a stability analysis of the system, assuming all planets lie on the same plane as planet c, demonstrated the planet is stable, consistent with the results of \cite{Bar04}. We performed this test with HNBody\footnote{HNBody is publicly available at http://janus.astro.umd.edu/HNBody/.} which accounts for any general relativistic precession of planet b. However, the amplitude of the \HD d signal is very small, $K\sim5 $\ms, at the level of expected stellar Doppler noise (Section~\ref{SR}). The higher-cadence HET data
set which could most clearly identify this object, does not span enough time for
an adequate fit to the longer period \HD c. An unconstrained fit of HET data alone
will not be possible for several more years. Therefore, we advise caution in the use of the elements listed
in Table~\ref{tab:allorb3} and in the adoption of this signal
as unequivocal evidence of a component d. Confirmation will require additional high-cadence RV observations, and/or future astrometry. A minimum mass component d would generate a peak to peak astrometric signature of 52 microarcseconds, likely detectable by Gaia \citep{Cas08}.

\section{Discussion and Conclusions}

\subsection{Discussion}
Given the adopted Table~\ref{tab:allorb3} errors, \HD c is either one of the most massive exoplanets or one of the least massive brown dwarfs.  We can compare our true mass to the (as of January 2010) 69 transiting exoplanetary systems, each also characterized by true mass, not \msini. As shown in the useful Exoplanetary Encyclopedia \citep{Sch09} only one companion (CoRoT-Exo-3 b, Deleuil \etal 2008) has a mass in excess of 13\mjupe.  Whether this 'brown dwarf desert' (Grether \& Lineweaver, 2006) \nocite{Gre06} in the transiting sample is due to the difficulty of migrating high-mass companions (bringing them in close enough to increase the probability of transit), or to inefficiencies in gravitational instability formation is unknown. 

The age of the host star, $\sim3.3$ By, would suggest that \HD c has not yet cooled to an equilibrium  temperature. An estimated temperature and self-luminosity for a 17\mjup object that is 3.3 By old can be found from the models of \cite{Hub02}. Those models predict that \HD c has an effective temperature, T$_{eff}\simeq400$K, and L=2.5e-7L$_{\sun}$, about 20 times brighter than what we estimated using these same models for \eps~b \citep{Ben06}. Unfortunately \HD~has about 16 times the intrinsic brightness of \eps, erasing any gain in contrast. We note that due to the eccentricity of the orbit, \HD c is actually within the present day habitable zone for a fraction of its orbit. As \HD ~continues to evolve and brighten, the habitable zone will move outward and \HD c will be in that zone for some period of time.

If the inner known companion \HD b is a minimum mass exoplanetary object (assuming $M = M sin i=0.8$\mjup), our 1 mas astrometric per-observation precision precludes detecting that 2 microsecond of arc signal.  Invoking (with no good reason) coplanarity with \HD c similarly leaves us unable to detect \HD b. However, with the motivation of our previous result for HD 33636 \citep{Bea07}, we can test whether or not \HD b is also stellar by establishing an upper limit from our astrometry. To produce a perturbation, $\alpha_b > 0.2$ mas (a 3-$\sigma$ detection, given $\sigma_{\alpha}=0.06$ mas from Table~\ref{tab:allorb3}), and the observed RV amplitude, K$_b$=59\ms, requires $M_b\sim0.1M_{\sun}$ in an orbit inclined by less than 0\fdg5. Our limit is  lower  than that established with the CHARA interferometer \citep{Bai08b}, who established a photometric upper limit of G5 V for the b component. While it might be possible to use 2MASS and SDSS  \citep{Ofe08} photometry of this object  to either confirm or eliminate a low-mass stellar companion by backing out a possible contribution from an M, L, or T dwarf, using their known photometric signatures \citep{Haw02, Cov07}, we lack precise (1\%) knowledge of the intrinsic photometric properties of a sub-giant star in the Hertzsprung gap with which to compare. 


\subsection{Conclusions}

	In summary, radial velocities from four sources, Lick and Keck  \citep{Wri09}, HJS/McDonald \citep{Wit09b},  and our new high-cadence series from the HET, were combined with \HST~astrometry to provide improved orbital parameters for \HD b and \HD c. Rotational modulation of star spots with a period P=31.66$\pm$0.17$^d$ produces 0.15\% photometric variations, spot coverage sufficient to produce the observed residual RV variations. Our simultaneous modeling of radial velocities and over three years of {\it HST} FGS astrometry yields  the signature  of a perturbation due to the outermost known companion, \HD c. Applying the Pourbaix \& Jorrisen constraint between astrometry and radial velocities, we obtain for the perturbing object \HD c a period, P = 2136.1 $\pm$ 0.3 d, inclination, $i$=48.3\fdg2 $\pm$ 3.7\arcdeg, and perturbation semimajor axis, 
$\alpha_c = 1.05\pm 0.06$ mas.  Assuming for \HD~a stellar mass $M_*$ = 1.48$\pm$0.05$M_{\sun}$, we obtain a mass for \HD~c, $M_b$ = 17.6$^{+1.5}_{-1.2}$\mjup,  within the brown dwarf domain. Our  independently determined parallax  agrees within the errors with \HIP, and  we find a close match in proper motion. Our HET radial velocities combined with others establish an upper limit of about one Saturn mass for possible companions in a dynamically stable range of companion-star separations, $0.2 \le a \le 1.2$ AU. RV residuals to a model incorporating components b and c contain a signal with an amplitude equal to the RMS variation with a period, P$\sim194^d$ and an inferred $a\sim0.75$ AU. While dynamical simulations do not rule out interpretation as a planetary mass companion, the low S/N of the signal argues for confirmation.

\acknowledgments

We thank the Carnegie-California Exoplanet Consortium, particularly J. T. Wright, for access to their improved \HD~velocities in advance of publication. Support for this work was provided by NASA through grants GO-10610, GO-10989, and GO-11210 from the Space Telescope 
Science Institute, which is operated
by the Association of Universities for Research in Astronomy, Inc., under
NASA contract NAS5-26555.  J.L.B. acknowledges support from the DFG 
through grants GRK 1351 and RE 1664/4-1, and the European CommissionÕs 
Seventh Framework Programme as an International Fellow (grant 
no.~PIFF-GA-2009-234866). The Hobby-Eberly Telescope (HET) is a joint project of the University of Texas at Austin, the Pennsylvania State University, Stanford University, Ludwig-Maximilians-Universit\"{a}t M\"{u}nchen, and Georg-August-Universit\"{a}t G\"{o}ttingen. The HET is named in honor of its principal benefactors, William P. Hobby and Robert E. Eberly. We thank the many Resident Astronomers and Telescope Operators whose efforts produced the high-quality spectra from which our HET velocities were extracted. T. H. was a Visiting Astronomer, Cerro Tololo Inter-American Observatory, National
Optical Astronomy Observatory, which is operated by the Association of
Universities for Research in Astronomy, Inc., under cooperative agreement
with the National Science Foundation.
This publication makes use of data products from the 
Two Micron All Sky Survey, which is a joint project of the University of 
Massachusetts 
and the Infrared Processing and Analysis Center/California Institute of 
Technology, 
funded by NASA and the NSF.  This research has made use of the SIMBAD database, 
operated at Centre Donnees Stellaires, Strasbourg, France; Aladin, developed and maintained at CDS; the NASA/IPAC Extragalactic Database (NED) 
which is operated by JPL, California Institute of Technology, under contract 
with 
NASA;  the Exoplanet Encyclopedia (grace \'{a} J. Schneider); and NASA's Astrophysics Data System Abstract Service. 
An anonymous referee motivated improvements to the clarity of presentation for which we are thankful.



\bibliography{/Active/myMaster.bib}

\clearpage

\begin{deluxetable}{lll}
\tablewidth{0in}
\tablecaption{\HD~Stellar Parameters \label{tbl-STAR}}
\tablehead{ 
\colhead{Parameter}& 
\colhead{Value}& 
\colhead{Source}
}
\startdata
SpT&G4 IV&1, 9\\
T$_{eff}$&5697 K&2\\
log g&3.94 $\pm$ 0.1&2\\
$[Fe/H]$&0.27 $\pm$ 0.05&2\\
age&3.28 $\pm$ 0.3 By&2\\
mass&1.48 $\pm$ 0.05 $M_{\sun}$&2\\
distance&40.0 $\pm$ 0.5 pc & 3\\
R&2.44 $\pm$ 0.22$ R_{\sun}$&4\\
v\,sin\,i&3.5 $\pm$ 0.5 \kms&5
\\V&5.90 $\pm$ 0.03&6\\
K&4.255 $\pm$ 0.03&7\\
V-K&1.65 $\pm$ 0.04&\\
i-z&0.06&8\\
g-r&0.55&8\\
r-i&0.15&8\\
\hline
\enddata
\\$^1$Fischer \etal 2001, $^2$\cite{TakY07} or \cite{Tak07},
$^3$parallax from Table~\ref{tbl-SUM},
$^4$\cite{Bai08a}, $^5$\cite{Val05}, $^6$SIMBAD,
$^7$2MASS, $^8$\cite{Ofe08}, $^9$\cite{Mur02}.
\end{deluxetable}

\begin{deluxetable}{lrccc}
\tablecaption{The RV Data Sets \label{tbl-RV}}
\tablewidth{0in}
\tablehead{ 
\colhead{Data Set}& 
\colhead{Coverage}& 
\colhead{Nobs} & 
\multicolumn{2}{c}{RMS [\ms]} }
\startdata
& & & 3C\tablenotemark{a} & 2C\tablenotemark{b}\\
\hline
Lick  & 1998.79-2008.22 & 109   & 10.34  &10.74 \\
HJS  &1995.72-1996.78 &  7  & 7.24 & 7.56 \\
Keck  &1996.92-2008.07 &  55   & 7.39 & 7.90 \\
HET  &2004.92-2008.98 &313\tablenotemark{c}  &5.75 & 5.92\\
\hline
&total&484& &
\enddata 
\tablenotetext{a}{solution including components b, c, and d.}
\tablenotetext{b}{solution including components b, c only.}\\
\tablenotetext{c}{reduced to 102 normal points.}
\end{deluxetable}

\begin{small}
\begin{deluxetable}{cccccc}
\tablewidth{2.75in}
\tablecaption{HET RV Data  \label{tbl-HETRV}}
\tablehead{ 
\colhead{JD-2450000}& 
\colhead{RV (m/s)}& 
\colhead{$\pm$ error} } 
\startdata
3341.779899	&	-105.27	&	7.77	\\
3341.898484	&	-118.43	&	7.25	\\
3355.845730	&	-102.05	&	7.34	\\
3357.859630	&	-105.27	&	7.48	\\
3358.724097	&	-87.82	&	7.11	\\
3359.729188	&	-82.07	&	8.70	\\
3360.849520	&	-65.85	&	7.80	\\
3365.817387	&	1.45	&	7.73	\\
3367.812640	&	-20.41	&	9.48	\\
3369.701315	&	-90.14	&	8.57	\\
3371.684761	&	-107.83	&	8.90	\\
3377.785833	&	-20.51	&	8.92	\\
3379.675805	&	-3.53	&	6.78	\\
3389.755622	&	-50.16	&	7.68	\\
3390.763879	&	-35.52	&	7.46	\\
3391.757235	&	-19.07	&	8.15	\\
3392.750785	&	-5.45	&	7.11	\\
3395.738803	&	2.78	&	6.97	\\
3414.693834	&	-103.63	&	10.96	\\
3416.683636	&	-74.10	&	8.61	\\
3665.892690	&	64.01	&	4.74	\\
3675.986919	&	6.80	&	5.07	\\
3676.846929	&	21.68	&	5.48	\\
3678.862452	&	42.80	&	5.63	\\
3681.843596	&	51.51	&	5.06	\\
3685.835951	&	-63.04	&	25.10	\\
3685.837515	&	-57.70	&	5.16	\\
3691.933851	&	27.17	&	5.07	\\
3692.834053	&	39.74	&	5.00	\\
3694.820769	&	63.79	&	5.48	\\
3695.817331	&	61.75	&	5.81	\\
3696.807526	&	47.54	&	4.91	\\
3697.813712	&	8.53	&	5.32	\\
3700.809061	&	-38.81	&	5.48	\\
3708.894418	&	64.36	&	6.38	\\
3709.886951	&	60.71	&	5.85	\\
3711.767580	&	20.98	&	5.68	\\
3712.875847	&	-26.85	&	7.10	\\
3724.839770	&	60.73	&	6.27	\\
3730.717661	&	-28.39	&	7.05	\\
3731.708724	&	-18.87	&	6.85	\\
3733.706335	&	16.30	&	6.90	\\
3735.713861	&	48.56	&	6.86	\\
3739.692156	&	48.28	&	6.40	\\
3742.684910	&	-46.27	&	5.94	\\
3751.775752	&	73.88	&	6.84	\\
3752.761925	&	74.25	&	6.79	\\
3753.773028	&	64.34	&	7.73	\\
3754.760139	&	35.42	&	6.79	\\
3755.751319	&	-9.88	&	6.38	\\
3757.639021	&	-39.51	&	6.53	\\
3758.754961	&	-29.76	&	6.22	\\
3764.745400	&	62.37	&	6.78	\\
3989.998171	&	88.56	&	5.09	\\
4020.924198	&	141.62	&	5.36	\\
4021.921835	&	158.20	&	5.52	\\
4022.926094	&	164.55	&	8.27	\\
4028.903039	&	75.06	&	5.75	\\
4031.882076	&	104.25	&	5.64	\\
4031.997118	&	111.67	&	5.86	\\
4035.887008	&	164.27	&	6.09	\\
4037.876174	&	185.49	&	5.35	\\
4039.869089	&	180.04	&	5.69	\\
4040.971815	&	145.86	&	5.30	\\
4043.860701	&	95.53	&	5.43	\\
4048.842947	&	150.59	&	5.30	\\
4048.939538	&	149.37	&	7.22	\\
4051.843854	&	187.96	&	5.64	\\
4052.839025	&	197.86	&	5.30	\\
4053.847354	&	193.90	&	6.22	\\
4054.832617	&	170.75	&	5.14	\\
4056.922854	&	86.05	&	5.72	\\
4060.915198	&	100.89	&	5.56	\\
4061.912707	&	128.08	&	5.65	\\
4062.807028	&	136.95	&	5.26	\\
4063.809422	&	157.36	&	4.87	\\
4071.890292	&	91.98	&	5.58	\\
4072.774474	&	89.84	&	6.22	\\
4073.892958	&	100.30	&	5.66	\\
4075.757176	&	129.10	&	5.57	\\
4105.804643	&	175.94	&	7.08	\\
4109.801302	&	221.86	&	6.78	\\
4110.690995	&	223.57	&	6.45	\\
4121.646747	&	210.00	&	7.33	\\
4128.729702	&	122.64	&	7.65	\\
4132.725529	&	163.95	&	10.45	\\
4133.718796	&	165.86	&	6.92	\\
4163.637004	&	202.53	&	6.48	\\
4373.962556	&	210.00	&	8.54	\\
4377.933380	&	277.65	&	4.60	\\
4398.878305	&	269.02	&	5.90	\\
4419.843012	&	243.10	&	5.81	\\
4424.821940	&	314.81	&	4.42	\\
4425.816251	&	307.27	&	6.65	\\
4475.769793	&	190.40	&	7.04	\\
4487.651953	&	153.73	&	6.93	\\
4503.606266	&	151.66	&	7.95	\\
4520.665744	&	175.50	&	6.58	\\
4726.967268	&	61.40	&	8.31	\\
4729.974635	&	-45.60	&	6.43	\\
4808.884683	&	8.79	&	7.32	\\
4822.725625	&	18.40	&	6.72	\\
\enddata
\end{deluxetable}
\end{small}


\begin{center}
\begin{deluxetable}{rlllrlll}
\tablewidth{0in}
\tablecaption{Log of \HD ~FGS Observations\label{tbl-LOOF}}
\tablehead{
\colhead{Epoch}&
\colhead{MJD\tablenotemark{a}}&
\colhead{Year}&
\colhead{Roll (\arcdeg)\tablenotemark{b}}
}
\startdata
1&53597.05445&2005.619588&285.709\\
2&53599.2528&2005.625607&284.644\\
3&53600.11884&2005.627978&284.236\\
4&53601.18477&2005.630896&283.74\\
5&53605.98201&2005.64403&281.607\\
6&53613.97829&2005.665923&280.001\\
7&53689.06154&2005.87149&244.998\\
8&53690.05794&2005.874217&244.998\\
9&53691.05421&2005.876945&244.998\\
10&53692.18891&2005.880052&244.998\\
11&53693.25725&2005.882977&244.998\\
12&53697.65138&2005.895007&244.998\\
13&53964.25536&2006.624929&284.764\\
14&53965.05198&2006.62711&284.386\\
15&54057.37329&2006.879872&244.998\\
16&54058.37467&2006.882614&244.998\\
17&54061.30276&2006.89063&244.998\\
18&54781.68145&2008.86292&250.063\\
19&54781.74804&2008.863102&250.063\\
20&54782.28072&2008.864561&250.063\\
21&54782.34731&2008.864743&250.063\\
22&54782.41389&2008.864925&250.063\\
23&54782.48048&2008.865107&250.063\\

\enddata
\tablenotetext{a}{MJD = JD - 2400000.5}\\
\tablenotetext{b}{Spacecraft roll as defined in Chapter 2, FGS Instrument 
Handbook \citep{Nel07}}
\end{deluxetable}
\end{center}

\begin{deluxetable}{rlllll}
\tablewidth{0in}
\tablecaption{ Astrometric Reference Stars\label{tbl-1}}
\tablehead{
\colhead{ID}
&\colhead{RA\tablenotemark{a}~~~ (2000.0)~}&
\colhead{Dec\tablenotemark{a}}&
\colhead{V\tablenotemark{b}}
}
\startdata
2&86.624482&1.182386&14.12\\
4&86.642054&1.194295&13.05\\
5&86.612529&1.183187&13.87\\
6&86.655567&1.157715&14.34\\
\enddata
\tablenotetext{a}{Positions from 2MASS.}
\tablenotetext{b}{Magnitudes from NMSU.}
\end{deluxetable}

\begin{deluxetable}{cccccc}
\tablewidth{0in}
\tablecaption{V and Near-IR Photometry \label{tbl-IR}}
\tablehead{\colhead{ID}&
\colhead{$V$} &
\colhead{$K$} &
\colhead{$(J-H)$} &
\colhead{$(J-K)$} &
\colhead{$(V-K)$} 
}
\startdata
1&5.9$\pm$0.03&4.255$\pm$0.036&0.58$\pm$0.24&0.73$\pm$0.23&1.65$\pm$0.05\\
2&14.12 0.03&11.88 0.021&0.44 0.03&0.49 0.03&2.24 0.04\\
4&13.05 0.03&11.239 0.024&0.29 0.04&0.35 0.04&1.81 0.04\\
5&13.87 0.03&10.303 0.023&0.75 0.03&0.90 0.03&3.57 0.04\\
6&14.34 0.03&10.567 0.021&0.70 0.03&0.93 0.03&3.77 0.04\\
\hline
\enddata
\end{deluxetable}

\begin{deluxetable}{ccccccc}
\tablewidth{0in}
\tablecaption{Astrometric Reference Star Adopted
Spectrophotometric Parallaxes \label{tbl-SPP}}
\tablehead{\colhead{ID}& \colhead{Sp. T.\tablenotemark{a}}&
\colhead{V} & \colhead{M$_V$} &\colhead{m-M}& \colhead{A$_V$}&
\colhead{$\pi_{abs}$(mas)}} 
\startdata
2&F2V&14.12&3&11.12&1.302&1.1$\pm$0.3\\
4&F0V&13.05&2.7&10.35&1.147&1.4 0.3\\
5&K0III&13.87&0.7&13.17&1.395&0.4 0.1\\
6&K1III&14.34&0.6&13.74&1.271&0.3 0.1\\
\enddata 
\tablenotetext{a}{Spectral types and luminosity class estimated from classification spectra, colors, and 
reduced 
proper motion diagram (Figures~\ref{fig-CCD} and~\ref{fig-RPM}).}\\
\end{deluxetable}

\begin{deluxetable}{cccccc}
\tablewidth{0in}
\tablecaption{\HD~and Reference Star Relative Positions\tablenotemark{a}   \label{tbl-POS}}
\tablecolumns{6} 
\centering
\tablehead{\colhead{Star}&\colhead{V}& \colhead{$\xi$ } & \colhead{$\sigma_{\xi}$} &  \colhead{$\eta$} &  \colhead{$\sigma_{\eta}$}}
\startdata
1\tablenotemark{b}&5.9& -2.55702&0.00013&730.32659&0.00022\\
2&14.1&57.28598&0.00016&661.31308&0.00016\\
4\tablenotemark{c}&13.05&-14.55002&0.00011&635.50294&0.00014\\
5&13.87&98.23526&0.00012&647.94778&0.00014\\
6&14.34&-29.22806&0.00012&775.06603&0.00014\\
\enddata
\tablenotetext{a}{~Units are arcseconds}
\tablenotetext{b}{epoch 2005.895 (J2000); constraint plate at  epoch 53965.039571, rolled to 284\fdg386 }
\tablenotetext{c}{RA = 86\fdg642054, Dec = +1\fdg194295, J2000}

\end{deluxetable}

\begin{deluxetable}{cccccc}
\tablewidth{0in}
\tablecaption{Final Reference Star Proper Motions \label{tbl-PM}}
\tablehead{
\colhead{ID}&
\colhead{V} &
\colhead{$\mu_\alpha$\tablenotemark{a}} &\colhead{$\sigma_{\mu_\alpha}$}&
\colhead{$\mu_\delta$\tablenotemark{a}}&\colhead{$\sigma_{\mu_\delta}$}}
\startdata 
2&14.12&-13.32&0.15&15.45&0.12\\
4&13.05&2.32&0.09&7.87&0.09\\
5&13.87&-12.26&0.11&13.13&0.09\\
6&14.34&3.76&0.10 &-4.91&0.09\\
\enddata
\tablenotetext{a}{$\mu_\alpha$ and $\mu_\delta$ are relative motions in mas
yr$^{-1}$ }
\end{deluxetable}


\begin{deluxetable}{ll}
\tablewidth{0in}
\tablecaption{Reference Frame Statistics, \HD~Parallax ($\pi$), and Proper Motion ($\mu_\alpha$,  $\mu_\delta$)\label{tbl-SUM}}
\tablewidth{0in}
\tablehead{\colhead{Parameter} &  \colhead{Value} }
\startdata
Study duration  &3.25 y  \\
number of observation sets    &   23  \\
reference star $\langle V\rangle$ &  13.85     \\
reference star $\langle (B-V) \rangle$ &1.1   \\
{\it HST}~Absolute $\pi$& 25.11 $\pm$ 0.19     mas \\
~~~~~~~Relative  $\mu_\alpha$& -78.69 $\pm$ 0.08 mas yr$^{-1}$\\
~~~~~~~Relative  $\mu_\delta$&  -141.96 $\pm$ 0.08  mas yr$^{-1}$\\
\\
\hline
{\it HIP 97} Absolute $\pi$ &23.57 $\pm$ 0.92 mas\\
~~~~~~~~~~Absolute  $\mu_\alpha$& -80.05 $\pm$ 0.81 mas yr$^{-1}$\\
~~~~~~~~~~Absolute  $\mu_\delta$& -141.79 $\pm$ 0.66 mas yr$^{-1}$\\
{\it HIP 07} Absolute $\pi$ &25.46 $\pm$ 0.4 mas\\
~~~~~~~~~~Absolute  $\mu_\alpha$& -79.12 $\pm$ 0.48 mas yr$^{-1}$\\
~~~~~~~~~~Absolute  $\mu_\delta$& -141.84 $\pm$ 0.35 mas yr$^{-1}$\\

\enddata
\end{deluxetable}

\begin{deluxetable}{lrrr}
\tablewidth{0in}
\tablecaption{HD38529: Orbital Parameters and Mass Including 'd' \label{tab:allorb3}}
\tablehead{\colhead{Parameter} & \colhead{b} & \colhead{c} &\colhead{d} }
\startdata
\multicolumn{4}{c}{RV}\\
\hline 
$K$ (\ms) & 58.63 $\pm$ 0.37 & 170.23 $\pm$ 0.41 & 4.83   $\pm$ 1.3\\
$\het~ \gamma$ (\ms) & 48.4 $\pm$0.6 & & \\
$HJS~ \gamma$ (\ms) & -4.7 $\pm$ 2.1  & & \\
$Lick~ \gamma$ (\ms) &-33.1 $\pm$ 0.7 & & \\
$Keck~ \gamma$ (\ms) & -85.2 $\pm$ 0.8& & \\
\hline
\multicolumn{4}{c}{Astrometry} \\
\hline     
$\alpha$ (mas) & & 1.05 $\pm$ 0.06 &  \\
$i$ (\arcdeg) & & 48.3 $\pm$ 3.7 &  \\
$\Omega$ (\arcdeg) & & 38.2 $\pm$ 7.7 &   \\
\hline
\multicolumn{4}{c}{Astrometry~and~RV} \\
\hline
$P$ (days)& 14.3103$\pm$ 0.0002&2136.14 $\pm$  0.29&193.9 $\pm$ 2.9\\
$T$\tablenotemark{a} (days) &50020.18 $\pm$ 0.08 & 47997.1 $\pm$ 5.9 &52578.5 $\pm$ 3.3\\
$e$  & 0.254$\pm$ 0.007 & 0.362 $\pm$ 0.002& 0.23  $\pm$ 0.13\\
$\omega$ (\arcdeg) &95.3$\pm$ 1.7 &22.1 $\pm$ 0.6 & 160 $\pm$ 9 \\
\hline
\multicolumn{4}{c}{Derived\tablenotemark{b} } \\
\hline 
$a$ (AU) & 0.131  $\pm$ 0.0015  & 3.697  $\pm$ 0.042&  0.75 $\pm$ 0.14  \\
$\alpha\sin{i}$ (AU) & 7.459e-05  $\pm$ 3.3e-07 & 3.116e-02  $\pm$ 6.21e-06 &  8.4e-05 $\pm$ 2.4e-05  \\
Mass func  (\msun)& 2.703e-10  $\pm$ 2.5e-10 & 8.85e-07 $\pm$ 5.0e-9 &  2.1e-12 $\pm$ 1.4e-12\\
$M\sin{i}\,(M_{J})$\tablenotemark{c}  & 0.90  $\pm$ 0.041 & 13.99   $\pm$ 0.59 & 0.17 $\pm$ 0.06 \\
$M\,(M_{J})$  & & {\bf 17.6  $^{+1.5}_{ -1.2}$} &  \\
\enddata
\tablenotetext{a}{\,T = T - 2400000.0}
\tablenotetext{b}{\,A  mass of 1.48 $\pm$ 0.05\msun \citep{Tak07}  for HD38529 was assumed.}
\tablenotetext{c}{\,The minimum mass.}
\end{deluxetable}

\begin{deluxetable}{lrrr}
\tablewidth{0in}
\tablecaption{HD38529: Orbital Parameters and Masses for a Two Component Fit. \label{tab:allorb2}}
\tablehead{\colhead{Parameter} & \colhead{b} & \colhead{c} }
\startdata
\multicolumn{3}{c}{RV}\\
\hline 
$K$ (\ms) & 59.17 $\pm$ 0.42 & 171.99 $\pm$ 0.59  \\
$\het~ \gamma$ (\ms) & 47.6 $\pm$0.7 &  \\
$HJS~ \gamma$ (\ms) & -4.9 $\pm$ 2.2  &  \\
$Lick~ \gamma$ (\ms) &-34.0 $\pm$ 0.8 &  \\
$Keck~ \gamma$ (\ms) & -86.9 $\pm$ 0.8&  \\
\hline
\multicolumn{3}{c}{Astrometry} \\
\hline     
$\alpha$ (mas) & & 1.05 $\pm$ 0.09   \\
$i$ (\arcdeg) & & 48.8 $\pm$ 4.0  \\
$\Omega$ (\arcdeg) & & 37.8 $\pm$ 8.2    \\
\hline
\multicolumn{3}{c}{Astrometry~and~RV} \\
\hline
$P$ (days)& 14.3104$\pm$ 0.0002&2134.76 $\pm$  0.40\\
$T$ (days) &50020.19 $\pm$ 0.08 & 48002.0 $\pm$ 6.2  \\
$e$  & 0.248$\pm$ 0.007 & 0.360 $\pm$ 0.003 \\
$\omega$ (\arcdeg) &95.9$\pm$ 1.7 &22.52 $\pm$ 0.7 \\
\hline
\multicolumn{3}{c}{Derived } \\
\hline 
$a$ (AU) & 0.131  $\pm$ 0.0015  & 3.695  $\pm$ 0.043 \\
$\alpha\sin{i}$ (AU) & 7.540e-05  $\pm$ 3.9e-07 & 3.149e-02  $\pm$ 7.37e-06   \\
Mass func  (\msun)& 2.792e-10  $\pm$ 4.4e-10 & 9.137e-07 $\pm$ 6.1e-9 \\
$M\sin{i}\,(M_{J})$ & 0.92  $\pm$ 0.043   & 14.13   $\pm$ 0.62  \\
$M\,(M_{J})$  & & {\bf 17.7  $^{+1.7}_{ -1.4}$ } \\
\enddata
\end{deluxetable}


%
%

\clearpage

\clearpage

\clearpage

\end{document}